\documentclass[12pt]{article}
\usepackage {amscd}
\usepackage{graphicx}
 \newcommand{\df}{\stackrel{\bigtriangleup}{=}}

\begin{document}

\author{Evgueni A. Haroutunian, Associate Member, IEEE}
\title{On the Shannon Cipher System With a Wiretapper Guessing Subject
to Distortion and  Reliability Requirements$^*$ \footnotetext{$^*$
The material in this paper was presented in part at the Third
Conference on Computer Science and Information Technologies,
Yerevan, Armenia, 2001, at the IEEE International Symposium on
Information Theory, Lausanne. Switzerland, 2002, and at the session
of NATO ASI ``Network Security and Intrusion Detection", Yerevan,
2005. The author is  with the Institute for Informatics and
Automation Problems of the Armenian National Academy of Sciences, 1
P. Sevak St.,  Yerevan 375014, Armenia. E-mail: evhar@ipia.sci.am.
}}
\date{}
\maketitle

{\it Abstract -} In this paper we discuss the processes in the
Shannon cipher system with discrete memoryless source and a guessing
wiretapper. The wiretapper observes a cryptogram of $N$-vector of
ciphered messages in the public channel and tries to guess
successively the vector of messages within given distortion level
$\Delta$ and small probability of error less than $\exp \{-NE\}$
with positive reliability index $E$. The security of the system is
measured by the expected number of guesses which wiretapper needs
for the approximate reconstruction of the vector of source messages.
The distortion, the reliability criteria and the possibility of
upper limiting the number of guesses extend the approach studied by
Merhav and Arikan. A single-letter characterization is given for the
region of pairs $(R_L,R)$ (of the rate $R_L$ of the maximum number
of guesses $L(N)$ and the rate $R$ of the average number of guesses)
in dependence on key rate $R_K$, distortion level $\Delta$ and
reliability $E$.

\vspace{3mm}


\vspace{5mm}
 \centerline{ \rm \sc I. Introduction} \vspace{2mm}

We investigate the procedure of wiretapper's guessing with respect
to fidelity and reliability criteria in the Shannon cipher system
 (see Fig. 1) \cite{sh1}.

\begin{center}
\unitlength=0.8mm \special{em:linewidth 0.4pt} \linethickness{0.4pt}
\begin{picture}(133.00,71.00)

\begin{scriptsize}
\put(104.00,47.00){\makebox(0,0)[cc]{${\bf x}$}}
\put(26.00,47.00){\makebox(0,0)[cc]{${\bf x}$}}
\put(26.00,27.00){\makebox(0,0)[cc]{${\bf u}$}}
\put(62.00,47.00){\makebox(0,0)[cc]{$w$}}
\put(23.00,25.00){\line(1,0){21.00}}
\put(55.00,45.00){\vector(1,0){24.00}}
\put(1.00,39.00){\framebox(22.00,12.00)[cc]{\scriptsize Source}}
\put(1.00,19.00){\framebox(22.00,12.00)[cc]{\scriptsize Key source}}
\put(23.00,45.00){\vector(1,0){10.00}}
\put(33.00,39.00){\framebox(22.00,12.00)[cc]{}}
\put(43.00,49.00){\makebox(0,0)[ct]{\scriptsize Cipherer}}
\put(43.00,41.00){\makebox(0,0)[cb]{$f_N$}}
\put(44.00,25.00){\vector(0,1){14.00}}
\put(79.00,39.00){\framebox(22.00,12.00)[cc]{}}
\put(90.00,49.00){\makebox(0,0)[ct]{\scriptsize Decipherer}}
\put(90.00,41.00){\makebox(0,0)[cb]{$f_N^{-1}$}}
\put(89.00,25.00){\vector(0,1){14.00}}
\put(44.00,25.00){\line(1,0){45.00}}
\put(101.00,45.00){\vector(1,0){10.00}}
\put(111.00,39.00){\framebox(22.00,12.00)[cc]{}}
\put(122.00,49.00){\makebox(0,0)[ct]{\scriptsize Legitimate}}
\put(122.00,41.00){\makebox(0,0)[cb]{\scriptsize receiver}}
\put(56.00,59.00){\framebox(22.00,12.00)[cc]{\scriptsize
Wiretapper}} \put(67.00,45.00){\vector(0,1){14.00}}
\end{scriptsize}

\end{picture}
\end{center}
\vspace{-20mm}
\begin{center}
\small Fig. 1. The Shannon cipher system with a guessing wiretapper.
\end{center}

 Encrypted vector of messages of a discrete
memoryless stationary source must be transmitted via a public
channel to a legitimate receiver. The key-vector is communicated to
encrypter and to decrypter by special secure channel protected
against wiretappers. After ciphering the vector of source messages
by a key-vector, the cryptogram is sent over a public channel to a
legitimate receiver, which can recover the original message  on the
base of the cryptogram and the same key-vector. A wiretapper that
eavesdrops a public channel aims to decrypt the source messages on
the base of cryptogram, within the framework of given distortion and
reliability, knowing the source statistics and the  encryption
function but not the key. The wiretapper makes sequential guesses
(suppositions), each time applying a testing mechanism by which he
can learn whether the estimate is successful (is within a given
distortion level). He stops if the answer is affirmative, or the
number of guesses attains the prescribed limit. The restriction of
the number of guesses is justified because it often happens that
when some time passes the task of guessing loses its actuality or
even the sense.

The guessing problem was first considered by Massey \cite{mas}, then
by Arikan \cite{arikan2} and recently by Malone and Sullivan
\cite{ghu}. The guessing subject to fidelity criterion was studied
by Arikan and Merhav in \cite{ar-mer1}, \cite{ar-mer3}, for
reliability criterion by Haroutunian and Ghazaryan in \cite{sb21},
for the Shannon cipher system with exact reconstruction of messages
by wiretapper by Merhav and Arikan in \cite{mer-ar} and by Hayashi
and Yamamoto in \cite{haya}. The Shannon cipher system with
wiretapper reconstructing source messages subject to fidelity
criterion was examined by Yamamoto in \cite{yam}. We study a
combination of these problems with additional reliability criterion
and restriction of the number of guesses by a limit $L(N)$ (less or
equal to the number of all messages in  ${\cal X}^N$). The Shannon's
rate-distortion concept generalization, introduced by Haroutunian
and Mekoush \cite{har-mek}, consists in studying the
rate-reliability-distortion dependence. We use the term {\it
reliability} instead of the longer term {\it error probability
exponent}. Applications of the reliability criterion ware
investigated for various multiterminal systems (see \cite{sb21},
\cite{sorrento} -- \cite{HHH}, \cite{Kan}, \cite{Tun}).

The security of the cipher system we measure by the expected number
of guesses needed for reconstruction of the source messages. That
approach was used also by Merhav and Arikan in \cite{mer-ar} and
earlier by Hellman in \cite{hell} and by Sgarro in \cite{sgr1},
\cite{sgr2}. But we characterize the activity of the system also by
the rate of the maximum number of wiretapper guesses, the distortion
level of the approximate reconstruction of messages and the value of
the reliability (exponent) $E$ in the upper estimate $\exp \{-NE\}$
of the probability of error of the wiretapper.

The objective of this paper is investigation of the optimal
correlations of noted characteristics of the described model.
Abstracts of results of the paper were published in  \cite{csit01},
\cite{esit02}.

\vspace{5mm} \centerline{ \rm \sc II. Definitions} \vspace{2mm}

We pass to detailed definitions. The discrete memoryless source is
defined as a sequence $\left\{ X_i\right\} _{i=1}^\infty$ of
discrete, independent, identically distributed (i.i.d.) random
variables (RVs) $X$ taking values in the finite set ${\cal X}$ of
messages $x$ of the source. Let
$$
P^{*}=\{P^{*}(x),\,x\in {\cal X}\}
$$
be the source messages generating probability distribution (PD)
which is supposed to be known also to the wiretapper.  Let ${\bf
X}=(X_1,X_2,\dots,X_N)$ be a random $N$-vector. Since we study the
memoryless source the probability of the vector  ${\bf
x}=(x_1,\dots,x_N)$, a realization of the random $N$-vector ${\bf
X}$, is
$$
P^{*N}({\bf x})=\prod\limits_{n=1}^NP^{*N}(x_n).
$$
The key-source $\{U\}$ is given by a sequence $\left\{ U_i\right\}
_{i=1}^\infty$ of binary  i.i.d. RVs, which take values from the set
${\cal U}=\{0,1\}$. The distribution $P_1^{*}=\{1/2,\,1/2\}$ is the
PD of the key bits. The key-vector ${\bf u}=(u_1, u_2, \dots, u_K)$
is a vector of $K$ bits and $P_1^{*K}({\bf u})=2^{-K}$. Let ${\bf
U}=(U_1,U_2,\dots,U_K)$ be a key-vector of $K$ binary RVs
 independent of the vector ${\bf X}$.

Denote by $\hat x$ values of RV $\hat X$ representing the wiretapper
reconstruction of the source message with values in the finite
wiretapper's reproduction alphabet $\hat {\cal X}$, in general
different from ${\cal X}$.

Correspondingly, by ${\cal X}^N$ and $\hat {\cal X}^N$ we denote the
$N$-th order Cartesian powers of the sets ${\cal X}$ and $\hat {\cal
X}$, by ${\cal U}^K$ -- the $K$-th order Cartesian power of the set
${\cal U}$.

We consider a single-letter distortion measure between source and
wiretapper reproduction messages:
$$
d:{\cal X}\times \hat{\cal X}\rightarrow \left[ 0;\infty \right).
$$
It is supposed that for every $x{\in}{\cal X}$ there exists at least
one $\hat x \in \hat {\cal X}$ such that $d(x,\hat x)=0$. The
distortion measure between a source vector ${\bf x}\in {\cal X}^N$
and a wiretapper reproduction vector $\hat{\bf x}=(\hat x_1,\hat
x_2,...,\hat x_N)\in \hat {\cal X}^N$ is defined as an average of
the corresponding component distortions:
\begin{equation}
\label{1} d({\bf x},\hat {\bf
x})=N^{-1}\sum\limits_{n=1}^Nd(x_n,\hat x_n).
\end{equation}

Let
$$
f_N:{\cal X}^N\times{\cal U}^K\rightarrow {\cal W} (N,K)
$$
be an encryption function with the set ${\cal W}(N,K)$ of all
possible for this $N$ and $K$ cryptograms ${w}$.  This function is
assumed to be invertible providing the key is given , i. e. there
exists the decryption function
$$
f_N^{-1}:{\cal W}(N,K)\times {\cal U}^K\rightarrow  {\cal X}^{N}.
$$
We denote by $W(N,K)$ the RV with values ${w}$. For each cryptogram
$w=f_N({\bf x}, {\bf u})$ the ordered list of sequential guesses of
the wiretapper
$$
{\cal G}_N(w)\df\{\hat {\bf x}_1(w),\hat {\bf x}_2(w),\dots , \hat
{\bf x}_{L(N)}(w)\} ,\,\hat {\bf x}_l(w)\in\hat {\cal X}^N,
\,l=1,2,\dots, L(N),
$$
with the {\it limit of the number of guesses} $L(N) \leq |\hat {\cal
X}|^N$, is called the {\it guessing strategy} of the wiretapper. For
a given guessing strategy ${\cal G}_N(w), w\in{\cal W}(N,K)$, we
name {\it guessing function} and denote by $G_N({\bf x}, w)$ the
function
$$
G_N:{\cal X}^N\times{\cal W}(N,K)\rightarrow \{1, 2, 3, \dots, L(N),
L(N)+1\},
$$
which shows index $l$ of the  first successful guessing vector $\hat
{\bf x}_l(w)\in {\cal G}_N(w)$, i. e. such minimal $l$ that $d({\bf
x}, \hat {\bf x}_l(w))\leq \Delta$. In other words $l$ is the
quantity of sequential guesses of the wiretapper  until the
successful estimate $\hat {\bf x}_l(w)$ of the source vector ${\bf
x}\in {\cal X}^N$ is found. $G_N({\bf x},w)$ equals $L(N)+1$ if the
guessing is stopped after $L(N)$ unsuccessful attempts.

For each distortion level $\Delta\geq 0$, a positive number $L(N)$
and a guessing strategy ${\cal G}_N(w)$ let us consider two sets of
vectors ${\bf x}$ of messages:

the first is the set of those ${\bf x}$ which can be successfully
deciphered by the wiretapper within $L(N)$ guessing attempts for
every key ${\bf u}$
$$
{\cal A}(w)\df{\cal A}(L(N),{\cal G}_N(w), \Delta)\df\{{\bf
x}:\,\forall {\bf u},\,\exists l\leq L(N),\,\, f_N({\bf x}, {\bf
u})=w, \,\, d({\bf x}, \hat {\bf x}_l(w))\leq \Delta\}
$$

$$
= \{{\bf x}:G_N({\bf x},w)\leq L(N)\},
$$
and the other with those ${\bf x}$, which can not be deciphered by
the wiretapper with necessary precision after $L(N)$ guesses
$$
\overline{ {\cal A}(w)}\df\{{\bf x}:\,\,\exists {\bf u}, \, \forall
l\leq L(N), \,\, f_N({\bf x}, {\bf u})=w,\,\,\, d({\bf x}, \hat {\bf
x}_l(w))> \Delta\}
$$
$$
 ={\cal X}^N-{\cal A}(w) =\{{\bf x}: \exists{\bf u},\, f_N({\bf x}, {\bf u})=w, \,G_N({\bf x},w)=L(N)+1\}.
$$
Respectively, the  probability of the wiretapper error (probability
of unsuccessful guessing) will be defined for each $w$ and $\Delta$
as
$$
e(L(N),{\cal G}_N(w), \Delta)\df 1-P^{*N}\left({\cal
A}(w)\right)=P^{*N}\left(\overline{{\cal A}(w)}\right).
$$
Just as in other problems of information theory
 \cite{HHH} we study
the exponential decrease by $N$ of the error probability with given
reliability (exponent) $E$. With $E\rightarrow 0$ we can obtain also
results corresponding to the case of error probability upper limited
by given small $\varepsilon>0$  not decreasing exponentially by $N$.

In this paper $\log$-s and $\exp$-s are taken to the base $2$.

Let $R_K$ be the {\it{key rate}}:
$$
R_K=N^{-1}\log 2^{K}=K/N.
$$

It is supposed that $L(N)$ also increases exponentially by $N$. The
guessing rates pair $R_L, R$ will be called (from the point of view
of cryptanalysis, i.e. the  wiretapper) $(R_K,E,\Delta)$-{\it
achievable} for given $E>0$, $\Delta\geq0$ and $R_K$, if for every
encryption function $f_N$ there exists a sequence of guessing
strategies ${\cal G}_N(w)$ such that
\begin{equation}
\label{1,1} \liminf\limits_{N \rightarrow \infty} N^{-1}\log
L(N)=R_L,
\end{equation}
\begin{equation}\label{2}
\liminf \limits_{N \rightarrow \infty} N^{-1}\log \mbox
{E}_{P^*,P_1^*} \{G_N({\bf X}, W)\}= R,
\end{equation}
and for all $w \in {\cal W}(N,K)$
\begin{equation}\label{3}
 e(L(N),{\cal G}_N(w), \Delta)\leq \exp \{-NE\}.
\end{equation}
Let us denote by ${\cal R}_{G}(P^*,R_K,E,\Delta)$ the set of all
$(R_K,E,\Delta)$-achievable (for wiretapper) pairs of guessing rates
$R_L, R$ and call it the {\it
guessingrates-keyrate-reliability-distortion region}. The boundary
of the region ${\cal R}_{G}(P^*,R_K,E,\Delta)$ we will designate by
$R_{G}(P^*,R_K,E,\Delta)$. It contains information on
interdependence of extremal values of rates $R$ and $R_L$, so it
will be convenient to conditionally name it {\it
guessingrate-keyrate-reliability-distortion function}.

The knowledge of such functional dependence is practically useful
because it gives possibility to ameliorate the security of the
cipher system by increasing of the key rate $R_K$, or by decreasing
of the number of allowed guesses $L(N)$.

In case  $E\rightarrow \infty$, ${\cal X}\equiv\hat{\cal X}$,
$\Delta = 0$, and $R_L=\log\left|{\cal X}\right|$
guessingrate-keyrate-reliability-distortion function becomes the
{\it guessingrate-keyrate function} $R_{G}(P^*,R_K)$ studied by
Merhav and Arikan in \cite{mer-ar}. A problem studied by Yamamoto in
the framework of the rate-distortion theory for Shannon cipher
system \cite{yam} corresponds to the case $L(N)=1$ with measuring of
the security of the system by the attainable minimum distortion.

Let $P=\{P(x),x\in{\cal X}\}$ be a PD on ${\cal X}$ and $Q=\{Q(\hat
x\mid x),\,x\in {\cal X},\,\hat x \in \hat{\cal X}\}$ be a
conditional PD on $\hat {\cal X}$ for given $x$, also we denote by
$PQ$ the marginal PD on $\hat{\cal X}:$
$$
PQ\df\{PQ({\hat x})=\sum\limits_xP(x)Q({\hat x}\mid x),\,\,{\hat
x}\in \hat {\cal X}\}.
$$

For given $x\in {\cal X}$ denote by $Q_P(\hat x\mid x)$ the
conditional PD on $\hat{\cal X}$  such that for each $\Delta$ the
following condition is fulfilled:
$
\mbox{E}_{P,Q_P}d(X,\hat X)\df\sum\limits_x P(x)Q_P(\hat x\mid
x)d(x,\hat x)\leq \Delta.
$

Let ${\cal M}(P,\Delta)$ be the set of all PDs $Q_P$ for given
$\Delta$ and $P$.

We use the following notations for entropy, information and
divergence:
$$
H_P(X)\df-\sum\limits_{x}P(x) \log P(x),
$$
$$
I_{P,Q}(X\wedge \hat X)\df\sum\limits_{x,\hat x}P(x)Q(\hat x\mid
x)\log\frac{Q(\hat x\mid x)}{\sum\limits _x P(x)Q(\hat x\mid x)},
$$
$$
D(P|| P^{*})\df\sum\limits_xP(x)\log\frac{P(x)}{P^*(x)}.
$$

For given $E>0$ consider the following set of PDs $P$ ``surrounding"
the generating PD $P^*$:
\begin{equation}
\label{alpha} \alpha \left( P^*,E\right)\df\{P:D(P|| P^{*})\leq E\}.
\end{equation}

We denote by $R(P,\Delta)$ the {\it rate-distortion function} for PD
$P$ (see \cite{berger}, \cite{cs-k}):
\begin{equation}
\label{rd1} R(P,\Delta)\df\min \limits_{Q_P \in {\cal M}(P,\Delta
)}I_{P,Q_P}(X\wedge \hat X),
\end{equation}
and by $R(P^*,E,\Delta)$ the {\it rate-reliability-distortion
function} (introduced in \cite{har-mek}): for source  with
generating PD of messages $P^*$
\begin{equation}
\label{rd2} R(P^*,E,\Delta)\df\max\limits_{P\in \alpha(P^*,E)}
R(P,\Delta).
\end{equation}
The first emergence of $R(P^*,E,\Delta)$ may be explained by Theorem
2 below. But we apply it to solving of the problem under
consideration.

In the next Section we formulate a theorem specifying the
guessingrates-keyrate-reliability-distortion region ${\cal
R}_{G}(P^*,R_K,E,\Delta)$. The proofs are exposed in Section IV.

\vspace{5mm}

\centerline{ \rm \sc III. Formulation of the Result}
\vspace{2mm}
The main result of the paper is the complete characterization of the
guessingrates-keyrate-reliability-distortion region ${\cal
R}_{G}(P^*,R_K,E,\Delta)$. We introduce the following region:

$\widetilde{{\cal R}}_{G}(P^*,R_K, E,\Delta)\df\{(R_L,R):$
\begin{equation}
\label{h1} \log|{\cal X}|\geq R_L\geq \min(R_K,R(P^*,E,\Delta)),
\end{equation}
\begin{equation}
\label{h2} R_L\geq R\geq\max\limits_{P\in \alpha \left( P^*,E\right)
}[\min(R_K,R(P,\Delta))-D(P|| P^{*})]\}.
\end{equation}

\begin{figure}[!ht]
\centering
\includegraphics[width=0.5\textwidth]{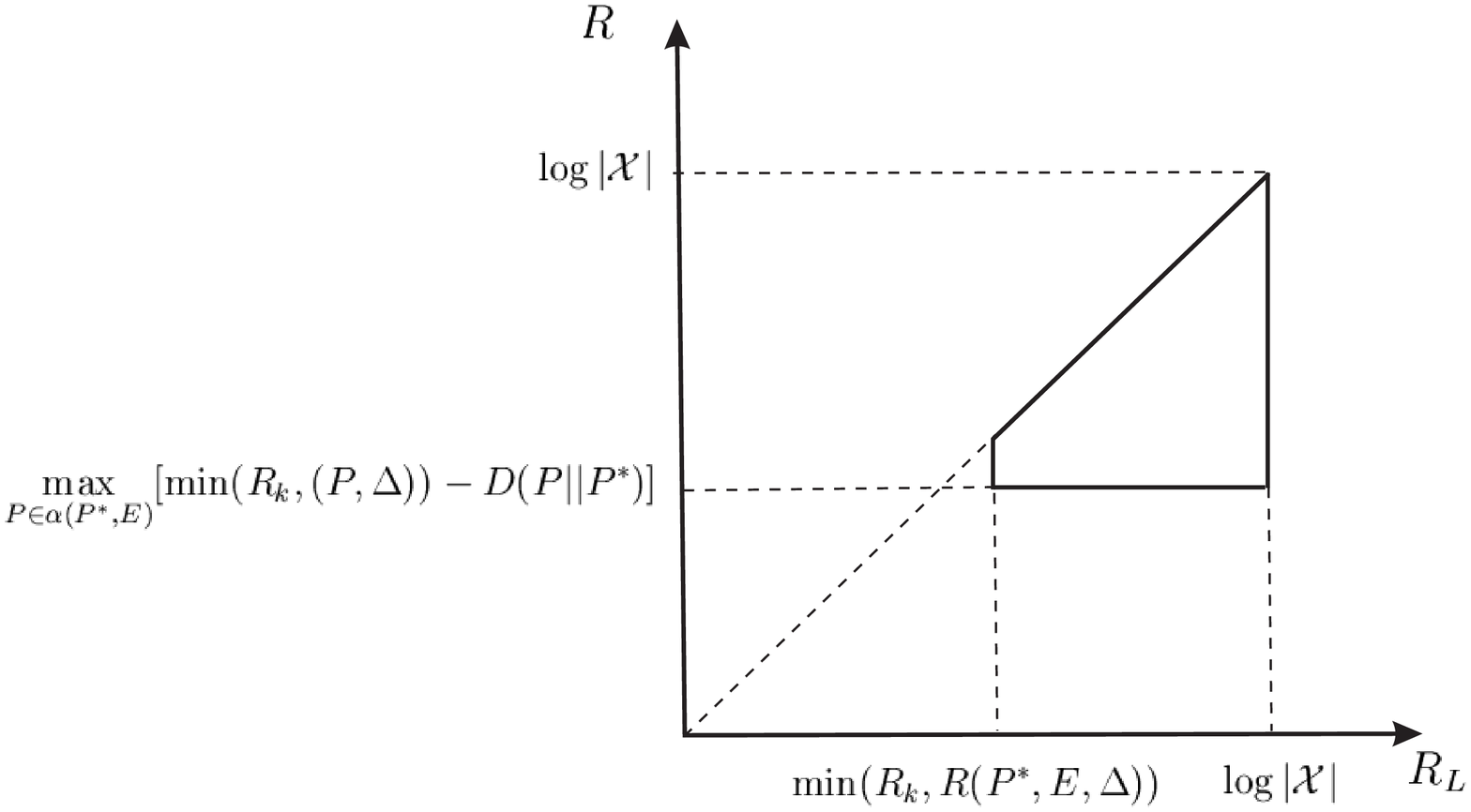}
%\caption{Illustration of ICI rule.}
% \label{fig:2}
\end{figure}

\centerline {\small Fig.2. Schematic diagram of region
$\widetilde{{\cal R}}_{G}(P^*,R_K, E,\Delta).$}

 \vspace{2mm}

 \vspace{2mm}
{\it Theorem 1:} For given PD $P^*$ on ${\cal X}$, every key rate
$R_K\geq 0$,  reliability $E>0$, and permissible distortion level
$\Delta \geq 0$,
\begin{equation}
\label{T1} {\cal R}_{G}(P^*,R_K,E,\Delta)=\widetilde{{\cal
R}}_{G}(P^*,R_K,E,\Delta).
\end{equation}

Theorem 1 comprises the following important particular cases. Denote
by $\widetilde{R}_{G}(P^*,R_K, E,\Delta)$ the boundary of the region
$\widetilde{{\cal R}}_{G}(P^*,R_K, E,\Delta)$.

\vspace{1mm}

 {\it Corollary 1:} When $E\to \infty$, and the strategy
permits the total exhaustion of the wiretapper reproduction vectors
set ($R_L=\log\left|{\cal X}\right|$) we get a solution of the
problem suggested by Merhav and Arikan \cite{mer-ar}, concerning the
reconstruction of the $N$-vector of messages by wiretapper within an
allowed level $\Delta$ of distortion  from the true vector
$$
\lim_{E\to \infty,\,\, R_L=\log\left|{\cal
X}\right|}R_{G}(P^*,R_K,E,\Delta)
$$
$$
=\lim_{E\to \infty,\,\,R_L=\log\left|{\cal
X}\right|}\widetilde{R}_{G}(P^*,R_K,E,\Delta)$$
$$
=\max\limits_{P}[\min(R_K,R(P,\Delta))-D(P|| P^{*})].
$$

\vspace{1mm}

{\it Corollary 2:} When $E\to \infty$, ${\cal X}\equiv\hat{\cal X}$,
$\Delta=0$, i.e. the wiretapper requires only the exact
reconstruction of sequences of source messages, and
$R_L=\log\left|{\cal X}\right|$, we arrive at the result of Merhav
and Arikan from \cite{mer-ar}:
$$
\lim_{E\rightarrow \infty,\,\, \Delta=0,\,\,R_L=\log\left|{\cal
X}\right|}R_{G}(P^*,R_K,E,\Delta)=\max
\limits_{P}[\min(R_K,H_P(X))-D(P|| P^ {*})].
$$
\vspace{1mm}

 {\it Corollary 3:} When $E\to 0$ we find that

$ \lim\limits_{E\rightarrow 0}\widetilde{{\cal
 R}}_{G}(P^*,R_K,E,\Delta)=\{(R_L,R):$
$$
R_L\geq\min(R_K,R(P^*,\Delta)),
$$
$$
 R\geq\min(R_K,R(P^*,\Delta))\}.
$$
This means that when the error probability decays by $N$ not
exponentially the maximal number of guesses may be greater than the
average number of guesses only by a factor which does not grow
exponentially by $N$.
\vspace{1mm}

  Explicit expressions of the
guessingrate-keyrate-reliability-distortion function
 for particular case of binary source and
Hamming distortion measure are presented together with some diagrams
in \cite{ASI05}.

\vspace{5mm}

\centerline{ \rm \sc IV. Proof of Theorem 1}

\vspace{2mm}

The first part of this Section will be appropriated to preliminary
necessary known results and tools. We apply the method of types (see
\cite{cov-thom}--\cite{cs}) in the proof of the theorem so let us
begin with the formulation of some basic concepts, notations and
relations of this method.

The type $P$ of a vector ${\bf x}\in {\cal X}^N$ is a PD
$P=\{P(x)=N(x|{\bf x})/N,\,x\in {\cal X}\}$, where $N(x|{\bf x})$ is
the number of repetitions of symbol $x$ among $x_1,\dots,x_N$. The
set of all PD-s $P$ on ${\cal X}$, which are types of vectors from
${\cal X}^N$ for given $N$, we denote by ${\cal P}({\cal X},N)$. The
set of vectors ${\bf x}$ of type $P$ will be denoted by ${\cal
T}_P^N(X)$ and also called the type.

Let $N(x,{\hat x}\mid {\bf x},{\hat{\bf x}})$ be the number of
repetitions of the pair $(x,{\hat x})$ in the pair of vectors $({\bf
x},{\hat{\bf x}})$. The conditional type of $\hat {\bf x}$ for given
${\bf x}$ from  ${\cal T}_P^N(X)$ is conditional PD $Q=\{Q(\hat
x|x),\, x\in {\cal X},\,\hat x\in \hat{\cal X}\}$ such that
$N(x,\hat x|{\bf x},\hat{\bf x})=N(x|{\bf x})Q(\hat x|x)=NP(x)Q(\hat
x|x)$ for $x\in {\cal X},\,\hat x\in \hat{\cal X}$. The set of all
vectors $\hat{\bf x}\in \hat{\cal X}^N$ of the conditional type $Q$
for given ${\bf x}\in {\cal T}_P^N(X)$ is denoted by ${\cal
T}_{P,Q}^N(\hat X|{\bf x})$. The set of possible conditional types
$Q$ for all ${\bf x}$ of the type $P$ is denoted by ${\cal
Q}(\hat{\cal X},P,N)$.

We use the following well known properties of types
(\cite{cov-thom}--\cite{cs}):
\begin{equation}
\label{type1} |{\cal P}({\cal X},N)|< (N+1)^{|{\cal X}|},
\end{equation}
and for each PD $P^{\prime}$ on $\cal X$

\begin{equation}
\label{type6} (N+1)^{-|{\cal X}|}\exp\{-ND(P|| P^{\prime})\}<
P^{\prime N}({\cal T}_P^N(X))\leq \exp\{-ND(P|| P^{\prime})\}.
\end{equation}

It turns out (as coming discussion shows) that the described
guessing problem is substantially interconnected with the problem of
source lossy coding  subject to distortion and reliability criteria.
The latter, according to \cite{har-mek}, as well as further works
\cite{ieee}, \cite{submi}, treats the Shannon rate-distortion coding
in view of the error probability exponential decay with exponent
$E$. This implies a more general optimal relation,
rate-reliability-distortion one $R(P^*,E,\Delta)$ between the coding
parameters instead of the rate-distortion function
$R(P^*,\Delta)$.

\begin{center}
\unitlength=0.8mm \special{em:linewidth 0.4pt} \linethickness{0.4pt}
\begin{picture}(133.00,53.00)

\begin{scriptsize}

\put(95.00,47.00){\makebox(0,0)[cc]{${\bf \widetilde x }$}}
\put(26.00,47.00){\makebox(0,0)[cc]{${\bf x}$}}
\put(63.00,47.00){\makebox(0,0)[cc]{$f_c(\bf x)$}}
\put(57.00,45.00){\vector(1,0){12.00}}
\put(1.00,39.00){\framebox(22.00,12.00)[cc]{\scriptsize Source}}
\put(23.00,45.00){\vector(1,0){12.00}}
\put(35.00,39.00){\framebox(22.00,12.00)[cc]{}}
\put(45.00,48.00){\makebox(0,0)[ct]{\scriptsize Encoder}}
\put(46.00,40.00){\makebox(0,0)[cb]{$f_c$}}
\put(69.00,39.00){\framebox(22.00,12.00)[cc]{}}
 \put(80.00,48.00){\makebox(0,0)[ct]{\scriptsize Decoder}}
\put(80.00,41.00){\makebox(0,0)[cb]{$g_c$}}
\put(91.00,45.00){\vector(1,0){12.00}}
\put(103.00,39.00){\framebox(22.00,12.00)[cc]{\scriptsize Receiver}}
\end{scriptsize}
\end{picture}
\vspace{-33mm}
\end {center}

\begin {center}
\small Fig. 3. The source lossy coding system.
\end {center}

\vspace{2mm}

For more details, let
$$
 f_c:{\cal X}^N \rightarrow \{1,2, \cdots, C(N) \}
$$
be an encoding mapping for source $N$-vectors with $C(N)$ standing
for the volume of the code. A backward mapping as a decoder of
source messages
$$
g_c: \{1,2, \cdots, C(N) \}  \rightarrow \hat{\cal X}^N
$$
is functioning with the encoder in a way to enable the probability
of error for $N$ large enough be restricted as follows:
\begin{equation}
\label{err} e(f_c,g_c,\Delta) \df \sum\limits_{{\bf x}\in {\cal
X}^N}P^{*N}\{{\bf x}: d \left({\bf x}, g_c(f_c({\bf x}))\right) >
\Delta \} \leq \exp \{-NE\},
\end {equation}
where $d \left({\bf x}, g_c(f_c({\bf x}))\right)$  is distortion
between transmitted source vector ${\bf x}$ and its reconstruction
$g_c(f_c({\bf x}))$. This distortion $d$ we supposed to be identical
to defined in (\ref {1})

For a predefined pair $\Delta \geq 0$ and $E > 0$ the
rate-reliability-distortion function $R(P^*,E,\Delta)$ specifies the
minimum achievable code rate $R \geq 0$ as a number to satisfy the
inequality $N ^{-1} \log C(N) \leq R +\varepsilon $ (where
$\varepsilon > 0$ is arbitrarily chosen beforehand) for every code
$(f_c,g_c)$, which validates (\ref {err}) kept $N$ appropriately
large.

The analytics for $R(P^*,E,\Delta)$ is given by the following
theorem -- a result constituting the inverse to the Marton's
exponent function from \cite{mar}.

\vspace{3mm}

 {\it Theorem 2} \cite{har-mek}:  For every $E>0$,
$\Delta \geq 0$ and $\varepsilon>0$, $\delta>0$ there exists a
sequence of such $N$-length block codes $(f_c,g_c)$ for source with
alphabet ${\cal X}$, generating PD $P^*$, and reproduction alphabet
$\hat{{\cal X}}$  that whenever $N \geq N_0(|{\cal
X}|,\varepsilon,\delta)$, then
$$
e(f_c,g_c,\Delta) \leq \exp\{-N(E+\delta)\}
$$
and
$$
N^{\ -1} \log\ C(N) \leq R(P^*,E,\Delta)+\varepsilon
$$
with $R(P^*,E,\Delta)$ defined in (\ref{rd1}), (\ref{rd2}).

Conversely, for every sequence of codes satisfying (\ref {err}) the
volume $C(N)$ cannot be too small:
$$
\liminf\limits_{N \rightarrow \infty} N^{\ -1} \log\ C(N) \geq
R(P^*,E,\Delta).
$$
\vspace{1mm}

 Theorem 2 is exposed with detailed proof in \cite{HHH}.
The derivation of Theorem 2 can be also observed from a more general
result in \cite{ieee} on robust descriptions system by eliminating
all the encoders except one. We only note here that the proof is
based on a random coding lemma about covering of types of vectors,
which is a modification of the covering lemmas from \cite{ahl},
\cite{ahl1}, \cite{cs-k}, \cite{sb21}, \cite{ieee}, \cite{IEEE2007}.

The proof of the following Proposition, which we have intention to
apply in solution of our guessing problem and which concerns with
coding of the vectors ${\bf x}$ of a separate type $P$ can
constitute the essential part of the proof of Theorem 2.

\vspace{3mm}

{\it Proposition:}  For each given type $P\in {\cal P}({\cal X},
N),$ every ${\bf x}\in {\cal T}_P^N(X), \, \Delta\geq0$, arbitrary
$\varepsilon>0$ and $N \geq N_0(P,\varepsilon)$ there exists a
sequence of such $N$-block codes $(f_{c,P},g_{c,P})$ of a volume
$C(P,N)$, that $d({\bf x},g_{c,P}(f_{c,P}(\bf x)))\leq \Delta\,$
with
$$
N^{\ -1} \log\ C(P,N) \leq R(P,\Delta)+\varepsilon,
$$
where $R(P,\Delta)$ is defined in (\ref{rd1}) and, conversely, for
every such code
$$
\liminf\limits_{N \rightarrow \infty} N^{\ -1} \log\ C(P,N) \geq
R(P,\Delta).
$$
\vspace{1mm}

We are ready now to proceed to the proof of Theorem 1. We intend to
prove that for every $R_K>0,\, E>0, \, \Delta>0$ the following
inclusions are valid
\begin{equation}
\label{type12} {\cal R}_{G}(P^*,R_K,E,\Delta)\supseteq
 \widetilde{\cal
R}_G(P^*,R_K,E,\Delta)\supseteq{\cal R}_G(P^*,R_K,E,\Delta),
 \end{equation}
 from where (\ref {T1}) follows.

\vspace{1mm}

The first inclusion in (13) is the converse kind statement from the
viewpoint of the security of the system and the direct statement
from the point of view of cryptanalysis. We have to prove that there
exists a guessing strategy the parameters $R_L, R$ of which meet
conditions (8) and (9).

 Now to prove the first inclusion in (\ref{type12}) consider a guessing strategy that
ignores the cryptogram. Represent ${\cal X}^N$ as a union of vectors
of various types
$$
{\cal X}^N=\bigcup\limits_{P\in {\cal P}({\cal X},N)}{\cal
T}_P^N(X).
$$

We frequently consider without additional mentioning PDs $P$ from
${\cal P}({\cal X},N)$, which are types for given $N$. When
$N\rightarrow\infty$ these types  converge to the corresponding
arbitrary PD-s from ${\cal P}({\cal X})$.

Based on the positive assertion of the Proposition independently of
a received $w$ the wiretapper can consider the collection of all
possible decoding vectors  as the guessing strategy for ${\bf x}\in
{\cal T}_P^N(X)$
$$
{\cal G}_N(w)=\{\hat {\bf x}_1(w),\hat {\bf x}_2(w),\dots , \hat
{\bf x}_{C(N,P)}(w)\}.
$$

Using  the right inequality in (\ref{type6}) and definition
(\ref{alpha}) of the set $\alpha (P^*,E)$  we can bound above the
probability of appearance of the source sequences of types $P$
beyond $\alpha(P^*,E+\delta)$ for some $\delta>0$ and $N$ large
enough as follows:
$$
P^{*N}(\bigcup\limits_{P\notin \alpha (P^*,E+\delta )} {\cal
T}_P^N(X))\leq (N+1)^{\left| {\cal X}\right| }\exp \{-N\min
\limits_{P\notin \alpha (P^*,E+\delta)}D(P|| P^{*})\}
$$
$$
\leq \exp \{-NE-N\delta +\left| {\cal X}\right| \log (N+1)\}\leq
\exp\{-NE \}.
$$
Therefore, to obtain the desired low level of $e(L(N),{\cal G}_N(w),
\Delta)$ it is sufficient that wiretapper constructs the guessing
strategy ${\cal G}_N(w)$ only for vectors of types $P$ from
$\alpha(P^*,E+\delta).$

We now pass to construction of such strategy. It is possible to
enumerate types $P$ from $\alpha (P^*,E+\delta)$ as
$P_1,\,P_2,\,\dots, P_{|\alpha (P^*,E+\delta)|}$ according to
nondecreasing values of corresponding rate-distortion functions
$R(P_i,\Delta)$ (for the sake of expressions simplicity we shall
write only $i$ instead of $P_i$ in $R(i,\Delta)$, ${\cal T}_i^N(X)$
and so on):
\begin{equation}
\label{133} R(1,\Delta)\leq R(2,\Delta)\leq\dots\leq
R(|\alpha(P^*,E+\delta)|,\Delta).
\end{equation}
We designate by $Q_{i}^{\mbox{\footnotesize min}}$ such conditional
PD from ${\cal M}(i,\Delta)$ that (see (\ref{rd1}) and (\ref{133}))
$$
C(i,N)=\exp\{N(\min \limits_{Q_i \in {\cal M}(i,\Delta )}I_{i,
Q_i}(X\wedge \hat
X)+\varepsilon)\}=\exp\{N(R(i,\Delta)+\varepsilon)\}.
$$
Let for fixed $i$ the set $\{\hat {\bf x}_{i,m} \in {\cal
T}_{i,Q_{i}^{\mbox{\scriptsize min}}}^N(\hat X),\,m=1,... ,C(i,N)\}$
be such a collection of decoding vectors that, according to the
Proposition, for $N$ large enough the set
$$
\{ {\bf x}: {\bf x} \in {\cal T}_{i,Q_{i}^{\mbox{\scriptsize
min}}}^N(X\mid \hat {\bf x}_{i,m}), \, f_{c,i}({\bf
x})=m,\,m=1,...,C(i,N)\},
$$
be a code for ${\cal T}_{i}^N(X)$. Let us consider the following
guessing strategy ignoring the cryptogram $w$:
$$
{\cal G}_N^*(w)\df\{\{\hat {\bf
x}_{1,m},\,m=1,...,C(1,N)\},...,\{\hat {\bf
x}_{L(N,P),m},\,m=1,...,C(L(N,P),N)\}.
$$

The number of required guesses $G_N^*({\bf x}, w)$ for ${\bf x}\in
{\cal T}_{i}^N(X)$, $P_i\in \alpha(P^*,E+\delta)$ and for each $w$
is upper bounded for $N$ large enough  (see (\ref{rd1}) and
(\ref{133}))
$$
G_N^*({\bf x}, w)\leq C(i,N)\leq\exp\{N(R(i,\Delta)+\varepsilon)\},
$$
and due to (\ref {rd2}) for every $\bf x$ of  type $P$ from
$\alpha(P^*,E+\delta)$ independently  of $w$ (independently of  $\bf
u$):
$$
G_N^*({\bf x}, w)\leq (N+1)^{|\cal
X|}\exp\{N(\max_{P_i\in\alpha(P^*,E+\delta)}R(i,\Delta)
+\varepsilon)\}\leq\exp\{N(R(P^*,E+\delta,\Delta)+2\varepsilon)\}.
$$

Sometimes, especially when $\Delta=0$, or $R_K$ is  small, it may be
appropriate for the wiretapper to carry out the key-search attack :
$$
{\cal G}_N^{**}(w)\df\{f^{-1}_N(w,{\bf u}_1), f^{-1}_N(w,{\bf
u}_2),\dots,f_N^{-1}(w,{\bf u}_{2^K}) \},
$$
where ${\bf u}_1$, ${\bf u}_2$, $\dots$, ${\bf u}_{2^K}$ is an
arbitrary numbering of all key-vectors of length $K$. Therefore, for
any given cryptogram $w$, the number of required guesses
$G_N^{**}({\bf x}, w)$ is upper bounded by the number of all
key-vectors
$$
G_N^{**}({\bf x}, w)\leq \exp K=\exp \{NR_K\}.
$$
This strategy gives to the wiretapper the exact $\hat {\bf x}={\bf
x}$ with the error probability equal to 0, but it remains to note
that for each ${\bf x}\in {\cal T}_P^N(X)$ when $R_K \geq
R(P,\Delta)$ there is no sense to guess key-vector ${\bf u}$. That
is why in that case the wiretapper may ignore $w$.

When $\exp\{{-K}\}>\exp\{{-N E }\}$ (the probability of each
possible key is greater than the desirable error probability) the
wiretapper has to test all $\exp K$ keys, that is in this case
$R_L=R_K$, and $E=\infty$. The  average rate $R$ is defined from the
equality
$$
R=\lim \limits_{N\rightarrow \infty}N^{-1}\log
[2^{-1}(\exp\{{NR_L}\}+1)].
$$
Thus, it follows that in the present instance
\begin{equation}
\label{cond5} R=R_L=R_K,
\end{equation}
hence (\ref {h1}), (\ref {h2}) and left inclusion in (\ref {type12})
are in force.

If
$$
 \exp\{{-NE}\}\geq\exp\{{-K}\}=\exp\{{-NR_K}\}
$$
the wiretapper can examine fewer than $\exp K$ keys. S/he can guess
successively with such rate of maximum number of guesses $R_L$ that
$$
\exp\{{NR_L}\}\exp\{{-NR_K}\}\geq 1-\exp\{{-NE}\}.
$$
Consequently for any small  $\varepsilon>0$ and sufficiently large
$N$
$$
\exp\{{NR_L}\}\geq\exp\{{NR_K}\}\{{1-\exp\{{-NE}\}\}}\geq\exp\{{N(R_{K}-\varepsilon)}\}.
$$
With the inequality  $R_K \geq R_L$, evident for the key searching,
we obtain that in this case again $R_L=R_K.$ But if the wiretapper
tests $\exp\{{NR_K}\}$ keys then the average number of guesses again
is equal to $2^{-1}(\exp\{{NR_K}\}+1)$. It means that (\ref {cond5})
is valid and (\ref {type12}) holds.

Combining these two guessing strategies as ${\cal G}_N^{***}(w)$,
when strategy ${\cal G}_N^{*}(w)$, or ${\cal G}_N^{**}(w)$ with the
least number of guesses is applied, we conclude that for a given
cryptogram $w$ the number of sequential wiretapper guesses for the
source vector ${\bf x}\in {\cal T}_{i}^N(X)$,
$P_i\in\alpha(P^*,E+\delta)$, for $N$ large enough is upper bounded
as follows
$$
G_N^{***}({\bf x}, w)\leq \min\{\exp
K,\exp\{N(R(i,\Delta)+\varepsilon)\}= \exp\{N\min\left(R_K,
R(i,\Delta)+\varepsilon\right)\}.
$$
Hence, for $N$ large enough, (see (\ref{rd2})) the required decrease
of error probability is attainable by the wiretapper if
$$
L(N)\leq\max\limits_{P \in \alpha(P^*, E+\delta)}\exp\{N\min( R_K,
R(i, \Delta)+\varepsilon)\}
$$
$$
=\exp\{N\min \left(R_K,R(P^*,E+\delta,\Delta)+\varepsilon\right)\}.
$$
Taking into account the independence of appearing of key-vectors and
source message vectors and using (\ref{type6}) and (\ref{type1}), we
can derive for $N$ large enough the upper estimate for the average
number of guesses:
\begin{eqnarray}
& & \mbox{E}_{P^*,P_1^*}\{G_N^{***}({\bf X}, W)\} \nonumber\\
& = & \sum\limits_{{\bf u}\in{\cal U}^K}P_1^{*K}({\bf u})
\sum\limits_{i:P_i\in \alpha(P^*,\,E+\delta)\cap{\cal P}({\cal X},
N)} \sum\limits_{{\bf x}\in {\cal T}_{i}^N(X)} P^{*N}({\bf
x})G_N^{***}({\bf x}, f_N({\bf x},{\bf u})) \nonumber\\
& \leq & \sum\limits_{{\bf u}\in{\cal U}^K} P_1^{*K}({\bf u})
\sum\limits_{P\in \alpha(P^*,\,E+\delta)\cap{\cal P}({\cal X},
N)}\sum \limits_{{\bf x}\in {\cal T}_{P}^N(X)} P^{*N}({\bf
x})\exp\{N\min \left(R_K,R(P,\Delta)+\varepsilon\right)\}\nonumber\\
& = &\sum\limits_{P\in \alpha(P^*,\,E+\delta)\cap{\cal P}({\cal X},
N)}\exp\{N\min
\left(R_K,R(P,\Delta)+\varepsilon\right)\}\sum\limits_{{\bf x}\in
{\cal T}_{P}^N(X)}P^{*N}({\bf x})\nonumber\\
& = &\sum\limits_{P\in \alpha(P^*,\,E+\delta)\cap{\cal P}({\cal X},
N)}\exp\{N\min
\left(R_K,R(P,\Delta)+\varepsilon\right)\}P^{*N}({\cal T}_{P}^N(X))\nonumber\\
& \leq & \sum\limits_{P\in \alpha(P^*,\,E+\delta)\cap{\cal P}({\cal
X}, N)}\exp\{N(-D(P|| P^{*})+\min
\left(R_K,R(P,\Delta)+\varepsilon\right))\}\nonumber\\
& \leq & \max \limits_{P\in \alpha \left(
P^*,\,E+\delta\right)}\exp\{N(-D(P|| P^{*})+ \min
\left(R_K,R(P,\Delta)+2\varepsilon\right))\}\nonumber\\
& = & \exp\{N\max \limits_{P\in \alpha \left(
P^*,\,E+\delta\right)}(-D(P|| P^{*})+ \min
\left(R_K,R(P,\Delta)+2\varepsilon\right))\}.\nonumber\
\end{eqnarray}

Therefore there exists a guessing strategy the rates of which
$R_L,R$ meet the inequalities
\begin{equation}
\label{rnum2} R_L\leq\min
\left(R_K,R(P^*,E+\delta,\Delta)+\varepsilon\right),
\end{equation}

\begin{equation}
\label {11} R\leq\max \limits_{P\in \alpha \left(
P^*,E+\delta\right) }(-D(P|| P^{*})+ \min
\left(R_K,R(P,\Delta)+2\varepsilon\right)).
\end{equation}
The pairs of values in right hand side correspond to the points in
region $\widetilde{{\cal R}}_G(P^*,R_K,E+\delta,\Delta)$, it means
that all points from $\widetilde{\cal R}_G(P^*,R_K,E+\delta,\Delta)$
will be $(R_K,E+\delta,\Delta)$-achievable for wiretapper as well.
Since $\varepsilon$ and $\delta $ can be made arbitrarily small and
all present expressions are continuous in $E$, we can consider
arbitrary PDs $P$ in (\ref{rnum2}) and (\ref{11}) and thus obtain
the left inclusion in (\ref{type12}).

\vspace{1mm}

Now we will prove the right inclusion in (\ref{type12})
$$
 \widetilde {\cal R}_G(P^*,R_K,E,\Delta)\supseteq{\cal
R}_{G}(P^*,R_K,E,\Delta).
$$
To prove this it is necessary to show that rates $R_L$ and $R$ of
every guessing strategy with keyrate $R_k$, reliability $E$, and
 distortion level $\Delta $ for arbitrary encryption algorithm must meet
 the right inequalities, correspondingly, in (\ref{h1}) and (\ref{h2}).
This is a converse statement from the point of view of
cryptographer.

It is supposed that the wiretapper knows algorithms of ciphering and
deciphering. We may assume also that the guesser knows the type $P$
of the source message ${\bf x}$, for such an informed guesser any
lower bounds on $L(N)$ and $\mbox{E}_{P^*,P_K^*}\{G_N^*({\bf X},
W)\}$  are lower bounds for uninformed guesser too.

 For each type $P$ the principal is the relation of two numbers:
$NR_K = K<NR(P,\Delta)$, or $K \geq NR(P,\Delta)$. In the first
occasion the key search is preferable for the wiretapper, in the
second situation s/he can guess ignoring the cryptogram. In fact the
wiretapper uses cryptogram $w$ only after guessing of key-vector
$\bf u$.

Let us start with the case
\begin{equation}
\label{cond1} R_K < R(P,\Delta).
\end{equation}

Denote by $\tilde{{\cal G}}_N(w,P)$  a guessing strategy  of the
wiretapper  that for any encryption function guarantees small error
probability: $e(L(N),\tilde{{\cal G}}_N(w,P), \Delta)\leq \exp \{-N
E\}$. Regardless  the source probability  distribution the optimal
guessing strategy  under the condition  (\ref {cond1}) is the
key-search attack. The wiretapper can then find the exact ${\bf x}$
applying description function $f^{-1}_N$ on the key vector and $w$.
Of course it is supposed that guessing of the exact $\bf {x}$ is
also acceptable for the wiretapper. We already know that in this
case the minimum values for $R$ and $R_L$ meet inequalities
 (\ref{h1}), (\ref{h2}).

Now let us consider the best strategy when  $P \in \alpha(P^*,
E+\delta)$ and
\begin{equation}
\label{cond2} \exp K \geq \exp\{NR(P,\Delta)\}.
\end{equation}

We also know that when $R_k \geq R(P,\Delta)$ the wiretapper can
guess each ${\bf x} \in {\cal T} _P^N(X)$ with distortion $\Delta$
and error probability less than $\exp \{-NE\}$ using less than $\exp
\{NR(P,\Delta)\}$ guesses, so key-search as demanding longer work is
not preferable. The question is: does another guessing strategy with
less than $\exp \{NR(P,\Delta)\}$ guesses exist? But every guessing
strategy $\{\hat {\bf x}_1(w),\hat {\bf x}_2(w),\dots , \hat {\bf
x}_{L(N,P)}(w)\}$ ignoring $w$ may be considered as a list for the
source encoding satisfying distortion and reliability criteria, so
according to the converse statement of the Proposition for $N$ large
enough $L(N,P)$ cannot be taken less than $\exp \{NR(P,\Delta)\}$.

Thus the numbers less than $\exp\{N\min (R_K, R(P,\Delta))\}$ cannot
be considered as limit $L(N,P)$, and for the common guessing
strategy inequality (\ref {h1}) is in force.

By averaging we obtain lower estimate for the expected number of
guesses:
\begin{eqnarray}
& & \mbox{E}_{P^*,P_1^*}\{G_N({\bf X},W \} \nonumber\\
& = & \mbox{E}_{P_1^*} \{ \mbox{E}_{P^*}\{G_N({\bf X},W )\} \}
\nonumber\\
& \geq & \sum\limits_{{\bf u} \in {\cal U}^K}P_1^{*K}({\bf u})
\sum\limits_{P\in \alpha(P^*,E+\delta)}\sum\limits_{{\bf x}
\in {\cal T}_{P}^{N}(X)}P^{*N}({\bf x})G_N({\bf x},w) \nonumber\\
& \geq & \sum\limits_{{\bf u} \in {\cal U}^K}P_1^{*K}({\bf u})
\sum\limits_{P\in \alpha(P^*,E+\delta)} \sum\limits_{{\bf x} \in
{\cal T}_{P}^{N}(X) \bigcap {\cal A}(w)}P^{*N}({\bf x})G_N({\bf x},w ) \nonumber\\
& = &  \sum\limits_{{\bf u} \in {\cal U}^K}P_1^{*K}({\bf u})
\sum\limits_{P\in \alpha(P^*,E+\delta)}P^{*N}({\cal A}(w))P^{*N}
\left(
{\cal T}_P^N(X)\right) \nonumber\\
&& \times \sum\limits_{l=1}^{\max\limits_{{\bf x} \in {\cal
T}_P^N(X)\bigcap {\cal A}(w)}G_N({\bf x},w)}l\Pr \{ \hat{\bf x}_l(w)
\mid {\bf x} \in {\cal T}_P^N(X) \bigcap {\cal A}(w) \}
\nonumber\\
& \geq & \sum\limits_{{\bf u} \in {\cal U}^K}P_1^{*K}({\bf
u})\sum\limits_{P\in \alpha(P^*,E+\delta)}(1-\exp\{-NE\})\exp\{-ND(P
\parallel P^*)\} \nonumber\\
&& \times \exp\{N(\min(R_K,\min\limits_{Q_P \in {\cal
M}(P,\Delta)}I_{P,Q_P}(X \wedge \hat{X})-\varepsilon))
\} \nonumber\\
& \geq & \exp\{N \max\limits_{P \in \alpha(P^*,E+\delta)}(\min(R_K,
R(P,\Delta)-D(P \parallel P^*)-2\varepsilon)) \}. \nonumber
\end{eqnarray}
In this calculation $P$ is type, but with growing of $N$ it
approaches arbitrary PD $P$. Hence for $N$ large enough
$$
R_L\geq N^{-1} \log L(N)-\varepsilon\geq
\min(R_K-\varepsilon,R(P^*,E+\delta,\Delta)-2\varepsilon),
$$
$$
R\geq N^{-1}\log \mbox{E}_{P^*P_{1}^*} \{G_N({\bf X},
W)\}-\varepsilon
$$
$$
\geq\max\limits_{P\in \alpha \left( P^*,
E+\delta\right)}(\min(R_K,R(P,\Delta))-D(P
\parallel P^*)-2\varepsilon).
$$
Granting arbitrariness of $\varepsilon$ and $\Delta$ we obtain
(\ref{h1}) and (\ref{h2}).

It rest to remark that comparison of cases (19) and (20) shows that
in condition (19) it is not possible to guess with $\Delta\neq0$ and
have smaller number of guesses, because approximate guessing will
need more than $\exp \{NR(P,\Delta)\}$ guesses, i. e. more than
$\exp \{NR_K\}$, which is enough for the exact reconstruction.

Therefore the proof  of the right inclusion in (\ref{type12}) is
completed.
\vspace{5mm}

\centerline{ \rm \sc Acknowledgment}

\vspace{2mm}

A. Ghazaryan, later L. Ghalechyan and recently A. Harutyunyan
participated in preparation of various versions of the paper. Author
would like to thank Prof. Maurer and the anonymous reviewers for
their valuable comments which helped to essentially improve the
exposition of the paper (it was submitted to the IEEE Transactions
on Information Theory for the  Special issue on Information
Theoretic Security).

\end{document}